\documentclass[prl,twocolumn,preprintnumbers]{revtex4}
\usepackage{graphicx,psfrag,dcolumn,bm,amsmath,amssymb}
\usepackage[dvips]{color}

\renewcommand{\vec}[1]{\boldsymbol{#1}}
\newcommand{\avg}[1]{\left\langle #1\right\rangle}
\newcommand{\mint}[4]{\int_{#2}^{#3}\!\!#1\,#4}
\newcommand{\fex}{\mathfrak f}
\newcommand{\Or}[1]{{\cal O}(#1)}

\newcounter{fnt}

\begin{document}

\title{Propagation and Relaxation of Tension in Stiff Polymers}

\author{Oskar Hallatschek$^1$} \author{Erwin Frey$^{1,2}$}
\author{Klaus Kroy$^{1,3}$} \affiliation{$^1$Hahn--Meitner Institut,
  Glienicker Stra\ss e 100, 14109 Berlin, Germany \\ $^2$Arnold
  Sommerfeld Center, Department of Physics,
  Ludwigs-Maximilians-Universit\"at M\"unchen, Theresienstra\"se 37,
  800333 M\"unchen, Germany \\ $^3$Institut f\"ur Theoretische Physik,
  Universit\"at Leipzig, Augustusplatz 10/11, 04109 Leipzig, Germany}

\date{\today}

\begin{abstract}
  We present a unified theory for the longitudinal dynamic response of
  a stiff polymer in solution to various external perturbations
  (mechanical excitations, hydrodynamic flows, electrical fields,
  temperature quenches \dots) that can be represented as sudden
  changes of ambient/boundary conditions. The theory relies on a
  comprehensive analysis of the non-equilibrium propagation and
  relaxation of backbone stresses in a wormlike chain.  We recover and
  substantially extend previous results based on heuristic arguments.
  New experimental implications are pointed out. (LMU-ASC 15/05)
\end{abstract}

\maketitle 

\begin{figure}[t]
\includegraphics[width=\columnwidth]{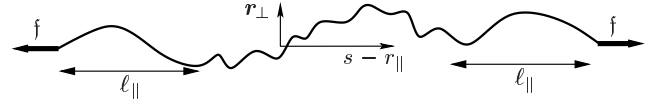}
\caption{\emph{Pulling} (schematic): In response to an external force
$\fex$, the thermally undulated contour $\vec r(s)=(\vec
r_{\!\perp},s-r_\|)^T$ is straightened within boundary layers of
growing width $\ell_\|(t)$.}
\label{fig:def}
\end{figure}

Despite considerable practical and interdisciplinary interest, it is
theoretically not yet fully understood how polymers respond to
external fields \cite{perkins-smith-chu:97,bohbot-raviv-etal:2004}.
Consider, e.g., the simple problem of an inextensible semiflexible
polymer suddenly stretched along its end-to-end vector by an external
force $\fex$ (\emph{Pulling}).  It was recognized before
\cite{seifert-wintz-nelson:96} that tension propagation (from the ends
into the bulk) is the key to understanding its dynamics: in response
to the spreading tension, the polymer stretches within a growing
boundary layer of length $\ell_\|$, as depicted in Fig.~\ref{fig:def}.
Depending on the setup, different tension propagation laws
$\ell_\|(t)$ have been
predicted~\cite{seifert-wintz-nelson:96,everaers-etal:99,%
  brochard-buguin-de_gennes:99,shankar-pasquali-morse:2002}.  In
particular, we contrast the above \emph{Pulling}-scenario with the
(inverse) \emph{Release}-scenario, where a constant pre-stretching
force $\fex$ is suddenly removed.  While for small $\fex$, one expects
$\ell_\|(t)\propto t^{1/8}$ in both cases \cite{everaers-etal:99}, the
predictions for strong force are markedly different:
$\ell_\|(t)\propto (\fex\, t)^{1/4}$ \cite{seifert-wintz-nelson:96}
for \emph{Pulling}, and $\ell_\|(t)\propto \fex^{3/4} t^{1/2}$
\cite{brochard-buguin-de_gennes:99} for \emph{Release}. However, the
precise meaning of ``strong'' and ``weak'', and the validity of the
diverse assumptions in
Refs.~\cite{seifert-wintz-nelson:96,brochard-buguin-de_gennes:99} are
not obvious. Here we develop from first principles a theory of stress
propagation and relaxation that allows us to unify and systematically
extend these previous results, and to derive (often analytically) the
longitudinal nonlinear response to various external perturbations.
After introducing the standard model of a semiflexible polymer, we
extend a heuristic argument of Ref.~\cite{everaers-etal:99} for
\emph{Pulling}. This elucidates the crossover from ``weak-'' to
``strong-force'' behavior and reveals the crucial length-scale
separation underlying our subsequent systematic analysis.

The wormlike chain model represents the polymer by a smooth
inextensible contour $\vec r(s,t)$ subject to the energy
\begin{equation}\label{eq:h}
{\cal H}=\frac\kappa2 \mint{ds}{0}{L}{\vec r''^2} +
\frac12\mint{ds}{0}{L}{f\vec r'^2}\;.
\end{equation}
The scalar force $f(s,t)$ (backbone ``tension'') is a Lagrange
multiplier function introduced \cite{goldstein-langer:95} to enforce
the inextensibility constraint $\vec r'^2=1$ for the tangent vector
$\vec r'\equiv \partial \vec r/\partial s$.  We require a bending
stiffness $\kappa$ such that the persistence length
$\ell_p=\kappa/k_BT$ is much larger than the contour length $L$, which
entails relative mean square transverse displacements of order
$\epsilon\equiv L/\ell_p\ll1$.  The elastic forces derived from $\cal
H$ have to balance thermal forces $\vec \xi$ (Gaussian white noise)
and Stokes friction, which (in the ``free-draining'' approximation)
enters through two local friction coefficients per unit length
$\zeta_\perp$, $\zeta_\|\approx \zeta_\perp/2$ for motion
perpendicular and parallel to $\vec r'$, respectively: $[\zeta_\| \vec
r'\vec r'+ \zeta_\perp(1- \vec r'\vec r')]\cdot\partial_t\vec
r=-\delta {\cal H}/\delta\vec r+ \vec \xi$. For the following, we
choose convenient units such that $\kappa \equiv \zeta_\perp \equiv
1$. Then, all dimensional quantities represent powers of length
(e.g.~$k_B T=\ell_p^{-1}$) and $\zeta_\|\equiv\zeta\approx1/2$.

We now turn to a heuristic discussion of \emph{Pulling} to leading
order in $\epsilon$. In the parameterization introduced in
Fig.~\ref{fig:def}, the exact equations of motion reduce to an
equation for the transverse displacements $\vec r_\perp$ alone,
\begin{equation}\label{eq:eom1}
\partial_t \vec r_{\!\perp} = -\vec r_{\!\perp}''''+f \vec r_{\!\perp}'' +
\vec\xi_{\!\perp} \;,
\end{equation}
with a spatially uniform tension $f=\fex\, \Theta(t)$ fixed by the
driving force at the boundaries; $\langle\vec\xi_{\!\perp}(s,t) \cdot
\vec\xi_{\!\perp}(s',t') \rangle=4 \ell_p^{-1} \delta(s-s')
\delta(t-t')$. The (higher order) longitudinal displacements $r_\|$
are slaved by the arclength constraint $r_\|'=\vec r_{\!\perp}'^2/2$.

From a simple scaling analysis of Eq.~(\ref{eq:eom1}), $\avg{\vec
  r_{\!\perp}}\!/t\approx \avg{\vec r_{\!\perp}}(\ell_{\!\perp}^{-4}+
\fex\, \ell_{\!\perp}^{-2})$, we deduce the characteristic dynamic
wavelength $\ell_{\!\perp}(t)$ corresponding to the (lowest) mode
equilibrated at time $t$. For instance,
$\ell_{\!\perp}(t_L^\perp)\equiv L$ defines the longest relaxation
time. Due to the competition between bending forces ($\propto
r_\perp\ell_{\!\perp}^{-4}$) and tension ($\propto r_\perp \fex\,
\ell_{\!\perp}^{-2}$), the growth of $\ell_{\!\perp}$ exhibits a
dynamic crossover from free relaxation ($\ell_{\!\perp}\propto
t^{1/4}$) to relaxation under tension ($\ell_{\!\perp}\propto
\sqrt{\fex\, t}$) at a characteristic time $t_\fex\equiv \fex^{-2}$
(Tab.~\ref{tab:pulling-growth-laws}/left).

By the above interpretation of $\ell_{\!\perp}$, the longitudinal
elongation of a subsection of arclength $\ell_{\!\perp}$ under a given
tension $\fex$ can be estimated by equilibrium theory. One has to
distinguish weak and strong tension relative to the internal
characteristic force scale $\ell_{\!\perp}^{-2}$ of the subsection,
which corresponds to the Euler buckling force of the subsection. For
\emph{weak} tension $\fex\ll \ell_{\!\perp}^{-2}$, the elongation
$\fex\, \ell_{\!\perp}^4 \ell_p^{-1}$ follows from linear response
\cite{mackintosh-kaes-janmey:95}. For \emph{strong} tension $\fex\gg
\ell_{\!\perp}^{-2}$ the subsection is virtually straight, so that the
elongation is equal to its equilibrium thermal contraction
$\ell_{\!\perp}^2\ell_p^{-1}$ caused by the bending undulations. Since
the whole polymer is subject to the same constant tension $\fex$, it
can be divided (at any time $t$) into $L/\ell_{\!\perp}(t)$
independent equilibrated subsections of length $\ell_{\!\perp}(t)$.
The total change $\Delta R(t)\equiv \lvert\avg{R(t)-R(0)}\rvert$ of
the end-to-end distance $R$ thus crosses over from $\Delta R\propto
L\fex\,\ell_{\!\perp}^{3} \ell_p^{-1}\propto t^{3/4}$ \cite{granek:97}
for $t\ll t_\fex$ to $\Delta R\propto L\ell_{\!\perp}
\ell_p^{-1}\propto t^{1/2}$ for $t\gg t_\fex$.

\begin{table}[t]
  \caption{The transverse equilibration length
    $\ell_{\!\perp}(t)$ and the tension propagation length
    $\ell_\|(t)$ both exhibit a crossover at $t_\fex\equiv\fex^{-2}$
    (here, for the \emph{Pulling} problem with $\fex\gg L^{-2}$, $t\ll
    t_L^\perp$).}
    \label{tab:pulling-growth-laws}
  \begin{ruledtabular}
    \begin{tabular}{r|cc|c} &
      $\ell_{\!\perp}(t)$  & & $\ell_\|(t)$ \qquad \quad \mbox{}\\ \hline
      $t\ll t_\fex$ &  $t^{1/4}$ &  & $t^{1/8} (\ell_p/ \zeta )^{1/2}$ 
     \hfill \cite{everaers-etal:99}\\
      $t\gg t_\fex$ &  $t^{1/2}\fex^{1/2}$ & & $t^{1/4}\fex^{1/4} 
      (\ell_p/\zeta)^{1/2}$ \quad  \cite{seifert-wintz-nelson:96}
    \end{tabular}
    \end{ruledtabular}
\end{table}

\begin{figure}[t]
 \includegraphics[width=\columnwidth]{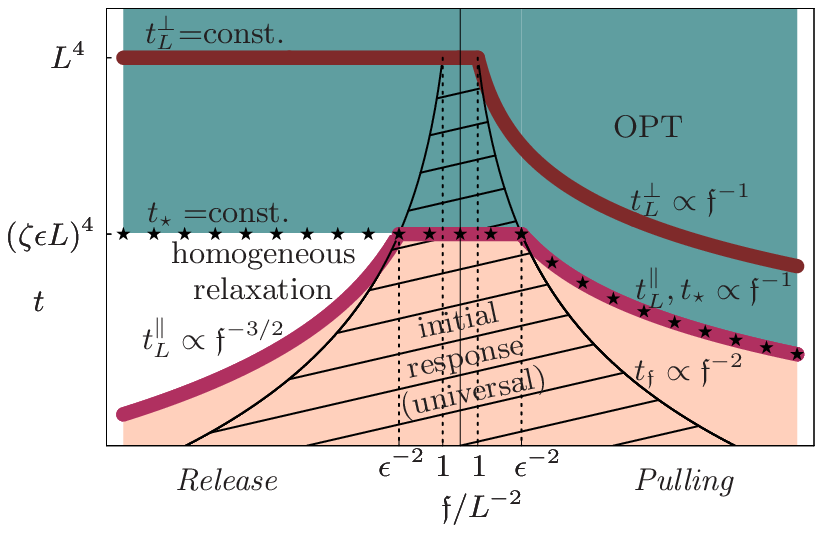}
\caption{Characteristic times (logarithmic scale) for \emph{Pulling} and
  \emph{Release} against the applied external force $\fex$ (linear
  scale). The time $t_\star$ (stars) separates regions where ordinary
  perturbation theory (OPT) applies (dark shaded) from regions (light
  shaded) of linear (hatched) and nonlinear tension propagation and
  from homogeneous tension relaxation (white). Whereas longitudinal
  friction is negligible for $t>t_\star$, it limits the dynamics for
  $t<t_\star$.}\label{fig:tf}
\end{figure}

These results comprise the predictions of ordinary perturbation theory
(OPT) to leading order. As evident from Eq.~(\ref{eq:eom1}), it neglects
longitudinal friction forces, which are of higher order in $\epsilon$.
However, the resulting $\Delta R$ obtained above implies a total
longitudinal friction $\zeta L \partial_t \Delta R$ crossing over from
$\zeta L^2 \fex\, t^{-1/4} \ell_p^{-1}$ to $\zeta
L^2(\fex/t)^{1/2}\ell_p^{-1}$ at $t_\fex$. Both expressions diverge
\cite{morse:98,everaers-etal:99} for $t\to 0$ indicating the breakdown
of OPT at short times. More precisely, for
times shorter than a certain $t_\star$~\footnote{$t_\star=(\zeta
  L^2\!/\ell_p)^4$ if $\fex \ll \ell_p^2/L^4$ and $t_\star=(\zeta
  L^2\!/\ell_p)^2 \fex^{-1}$ if $\fex\gg\ell_p^2/L^4$. See
  Fig.~\ref{fig:tf}.} the longitudinal friction would exceed the
driving force $\fex$. This was recognized as a consequence of the
omission of tension propagation in Eq.~(\ref{eq:eom1}): it was argued
\cite{everaers-etal:99} that actually only segments up to a distance
$\ell_\|(t)$~\footnote{Identifying $\ell_{\perp,\|}$ as correlation
  length of $\partial_t r_{\perp,\|}$, respectively, justifies the
  notation.}  from the ends are set into longitudinal motion causing
longitudinal friction. The proper expression for the total
longitudinal friction thus follows from the above upon substituting
$L$ by $\ell_\|$. The balance of longitudinal friction and external
force can now be satisfied by choosing the size $\ell_\|(t)$ of the
boundary layer according to Tab.~\ref{tab:pulling-growth-laws}/right.
Hence, the putative ``weak- and strong- force'' cases
$\ell_\|\propto t^{1/8}$ \cite{everaers-etal:99} and $\ell_\|\propto
t^{1/4}$ \cite{seifert-wintz-nelson:96} are identified as asymptotes of
a ``short-long time'' crossover (still) signaling the change from
``free'' to ``forced'' relaxation at $t=t_\fex$.

In summary, the foregoing discussion reveals: (\emph{i}) the breakdown
of OPT at times $t<t_\star$, where
(\emph{ii}) the neglected longitudinal friction actually limits the
relaxation to boundary layers of size $\ell_\|$; (\emph{iii}) the
crossover from free to forced relaxation at $t=t_\fex$; (\emph{iv})
the \emph{scale separation} $\ell_{\!\perp}/\ell_\| \propto
\epsilon^{1/2}\ll 1$ (Tab.~\ref{tab:pulling-growth-laws}).

The latter lends itself as starting point for a \emph{multiple-scale}
calculus to separate the physics on different dynamic scales and
obtain an improved (``multiple-scale'') perturbation theory (MSPT)
that is regular in the limit $t\to0$ while $\epsilon\ll1$ is fixed.
The procedure is similar to that for athermal dynamics
\cite{hallatschek-frey-kroy:04} and will be documented in detail
elsewhere \cite{hallatschek-etal:tbp}. The basic idea is to regard
functions $g(s)$ as functions $g(s,\bar s\epsilon^{1/2})$ of rapidly
and slowly varying arclength parameters $s$ and $\bar s
\epsilon^{1/2}$ that are treated as independent variables.  Closed
equations for the dynamics on the scale $\bar s \epsilon^{1/2}$ are
obtained upon averaging $\bar g(\bar s)\equiv
\mint{ds}{l}{\phantom{l}}{g(s,\bar s \epsilon^{1/2})/l}$ over the
fluctuations on the microscale ($\ell_{\!\perp}\ll l \ll \ell_\|$).
To leading order, we get \cite{hallatschek-etal:tbp} $f= \bar f(\bar
s,t)$,
\begin{equation}\label{eq:eom2}
  \partial_t \vec r_{\!\perp} = -\vec r_{\!\perp}''''+ \bar f\vec
  r_{\!\perp}'' + \vec \xi_{\!\perp} \;, \quad \text{and} \quad
  \partial_{\bar s}^2\bar f = - \zeta \overline {\partial_t r_\|'} \;.
\end{equation}
This provides the sought-after rigorous local generalization of the
above, heuristically renormalized force balance.  The arclength
average extending over many ($l/\ell_{\!\perp}\gg1$) uncorrelated
sections of length $\ell_{\!\perp}$ subject to the same tension $\bar
f(\bar s)$ can be interpreted as a coarse-graining that
\emph{effectively generates a local ensemble average}:
$\overline{\partial_t r_\|'} \sim \langle \partial_t r_\|'\rangle$ for
$\epsilon\to0$. Only the ``systematic'' $\Or1-$variations of the
tension are retained, while its $\Or\epsilon-$noise is leveled out, so
that the longitudinal Eq.~(\ref{eq:eom2}) represents
\emph{deterministic} dynamics: local longitudinal motion is driven by
tension gradients (like in a thread pulled through a viscous medium).

Integrating the longitudinal Eq.~(\ref{eq:eom2}) over time expresses
the change of the thermal contraction $r_\|$ in terms of the
time-integrated tension $F \equiv \mint{d t'}{0}{t}{f}$, namely
$\bigl\langle r_\|'(\bar s,t)-r_\|'(\bar s,
0)\bigr\rangle=-\partial_{\bar s}^2\bar F/\zeta$. Since $2r_\|'= \vec
r_{\!\perp}'^2$ from the arc\-length constraint, we integrate the
transverse Eq.~(\ref{eq:eom2}) for $\langle\vec
r_{\!\perp}'^2\rangle(\bar s,t)=
{\bigl\langle\bigl[\frac1L\sum_q\mint{dt'}{-\infty}{t}{q
    \chi_{\!\perp}(q;t,t')\vec\xi_{\!\perp}(q,t')}\bigr]^2\bigr\rangle}$.
Here
\begin{equation}
  \label{eq:tszbility}
  \chi_{\!\perp}(q;t,t')\equiv e^{ - q^4(t-t')-q^2[\bar
    F(\bar s, t)-\bar F(\bar s,t')]}
\end{equation}
should be recognized as the susceptibility for the response of the
Fourier modes of $\vec r_{\! \perp}$ to transverse forces.  Note that
the $\bar s-$dependence of $\langle\vec r_{\!\perp}'^2\rangle$ is
purely adiabatic, as it is parametrically inherited from $\bar F(\bar
s,t)$. Altogether, Eq.~(\ref{eq:eom2}) is condensed into a single
equation for $\bar F$~\footnote{The right-hand side of this equation
  has recently independently been derived in a related context
  \cite{bohbot-raviv-etal:2004}.},
\begin{equation}\label{eq:master}
 \frac{\partial_{\bar s}^2\bar
 F}\zeta =\mint{\frac{dq}{\pi}}{0}{\infty}{\left[
 \frac{1-\chi_{\!\perp}^2(q;t,0)}{\ell_p^-(q^2+f^-)} -\frac{2
 q^2}{\ell_p}\mint{dt'}{0}{t}\chi^2_\perp(q;t,t')\right]}\;.
\end{equation}
Indices ``$-$'' referring to $t<0$ were introduced to allow the system
to be prepared in equilibrium with ambient/boundary conditions
different from those prescribed for $t\geq0$.  Taking $\ell_p$ (for
$t\geq0$) to $\infty$, only the first term in the integrand remains,
which thus accounts for the deterministic relaxation of the initial
thermal contraction (set by $\ell_p^-$, $f^-$).  In this limit, the
``zero-temperature'' buckling dynamics analyzed in
Ref.~\cite{hallatschek-frey-kroy:04} is recovered. For finite
$\ell_p$ the second term in the brackets represents the action of the
thermal forces for $t>0$, which aim to establish the equilibrium
contraction.

To further unravel the physical implications of Eq.~(\ref{eq:master}),
we begin with the tension propagation regime $\ell_\|\ll L$, where $L$
is irrelevant. Problems like \emph{Pulling} and \emph{Release} still
depend on four length scales ($\ell_p^-=\ell_p, \fex^{-1/2},s ,
t^{1/4}$). Yet, Eq.~(\ref{eq:master}) is \emph{solved exactly}
by the two-variable scaling ansatz
\begin{equation}\label{eq:xover}
f(\bar s,t)= \fex \; \Phi(\bar s/s_\fex, t/t_\fex) \;.
\end{equation}
With $t_\fex\equiv \fex^{-2}$ and $s_\fex\equiv
(\ell_p/\zeta)^{1/2}t_\fex^{1/8}$ Eq.~(\ref{eq:xover}) can be shown to
eliminate the parameter dependencies in Eq.~(\ref{eq:master}) and in
its boundary conditions.  The scaling function $\Phi$ describes how
sudden changes of the tension at the ends spread into the bulk of the
polymer.  In the limits $t \ll t_\fex$ and $t\gg t_\fex$
Eq.~(\ref{eq:xover}) reduces to the simple (one-variable) scaling
laws $\Phi\sim (t/t_\fex)^\alpha \phi (\bar s/\ell_\|)$ with
the tension propagation length $\ell_\|\equiv s_\fex\,(t/t_\fex)^z$.
Note that this crossover scenario, as well as the expressions for
$t_\fex$, $s_\fex\approx\ell_\|(t_\fex)$, and $\ell_\|$ are consistent
with our above heuristic observations for \emph{Pulling}
(Tab.~\ref{tab:pulling-growth-laws}). In fact, this structure
generally emerges for all problems involving a single external force
scale. It is implicitly understood that $ \phi$, $\alpha$, and
$z$ will generally not only depend on the kind of external
perturbation applied, but will also be different in both scaling
limits $t/t_\fex\to 0,\infty$. In the following, these limits are
analyzed in more detail.

For $t\ll t_\fex$ Eq.~(\ref{eq:master}) may be linearized in $f$ and
the scaling function $ \phi$ can be obtained analytically
\cite{hallatschek-etal:tbp}. In contrast to $ \phi$, the
corresponding exponent $z=1/8$ does not depend on the boundary
conditions. It already follows from requiring $\phi$ to become
$\fex-$independent, as in linear response. The short-time dynamics
for strong external force is thus closely related to the linear
response. As established by our heuristic discussion of
\emph{Pulling}, this is due to the relaxation of subsections with
Euler forces $\ell_{\!\perp}^{-2}$ much larger than the external
force.  Note, however, that the limit $\fex\to0$ is problematic, as it
does not interchange with $\epsilon\to0$~\footnote{Our
    identification of arclength averages with (local) ensemble
    averages after Eq.~(\ref{eq:eom2}) breaks down for
    $\fex<(\zeta/\ell_p)^{1/4}t^{-7/16}$, where fluctuations in the
    tension become comparable to its average
    \cite{hallatschek-etal:tbp}.}. In fact, extending
Eq.~(\ref{eq:master}) to linear response amounts to an uncontrolled
factorization approximation $\avg{f r_{\!\perp}^2}\to
\avg{f}\avg{r_{\!\perp}^2}$.

\begin{figure}[t]
 \includegraphics[width=\columnwidth]{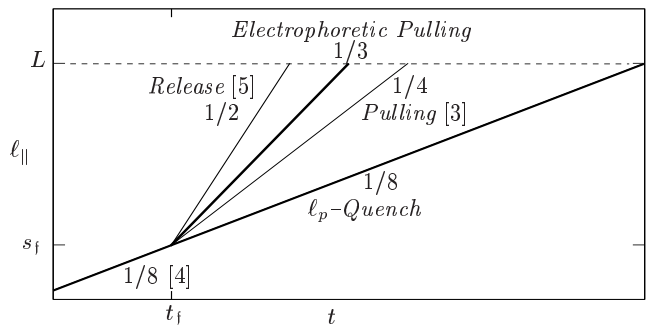}
  \caption{Double-logarithmic sketch of the tension propagation laws
    $\ell_\|(t)\propto t^z$.  At $t_\fex=\fex^{-2}$ they cross over
    from a universal short-time regime to (problem-specific)
    tension-dominated intermediate asymptotics, except for weak
    forces, $\fex < \ell_p^2/L^4$, and for
    $\ell_p-$\emph{Quenches}. The propagation ends when
    $\ell_\|(t)\approx L$.} \label{fig:lf}
\end{figure}

For $t\gg t_\fex$ the dynamics becomes nonlinear in the external force
and starts to depend on the force protocol.  Previously predicted
power laws can be recovered from Eq.~(\ref{eq:master}) by employing
different approximations to its right hand side.  In the
\emph{taut-string approximation} of
Ref.~\cite{seifert-wintz-nelson:96} one neglects for $t>0$ bending and
thermal forces against the tension, i.e.\ one drops the $q^4-$term in
the expression Eq.~(\ref{eq:tszbility}) for $\chi_\perp$ and sets
$\ell_p\to\infty$. The complementary \emph{quasi-static
  approximation} of Ref.~\cite{brochard-buguin-de_gennes:99} amounts
to the omission of memory effects, i.e.\ to the assumption of
instantaneous equilibration of tension and stored length,
$F(t)-F(t')\to f(t)(t-t')$. A careful analysis of
Eq.~(\ref{eq:master}) \cite{hallatschek-etal:tbp} shows that either of
these approximations becomes rigorous in the intermediate asymptotic
regime defined by $t\gg t_\fex$, $\ell_\|\ll L$.  The ``pure''
\footnote{We postpone ``mixed'' scenarios (e.g.\ strong stretching of
  a pre-stretched polymer) that involve more than one external force
  scale.  Contrary to previous expectations
  \cite{seifert-wintz-nelson:96,brochard-buguin-de_gennes:99}, they
  lead to multiple crossovers even for $t\gg t_\fex$.}  scenarios are
summarized in Fig.~\ref{fig:lf} and for the cases \emph{Pulling} and
\emph{Release} also in Fig.~\ref{fig:tf} and Tab.~\ref{tab:tf}.

\setcounter{fnt}{\value{footnote}}

In Fig.~\ref{fig:lf} we have moreover displayed results for sudden
changes in persistence length from $\ell_p^-$ to $\ell_p\neq\ell_p^-$
(``$\ell_p-$\emph{Quench}'') and \emph{Electrophoretic Pulling}, which
have not been discussed before.  The second is a variant of the
\emph{Pulling} problem, where the external force is applied along the
whole contour of an end-grafted polymer as is the case for a
hydrodynamic flow or an electric field.  The $\ell_p$-\emph{Quench}
is exceptional in that there is \emph{no external force scale}
involved, so that Eq.~(\ref{eq:master}) can be solved by a simple
one-variable scaling ansatz $f(\bar s,t)=t^{-1/2}\varphi (\bar
s/\ell_\|)$ with $\ell_\| \approx (\ell_p/\zeta)^{1/2}t^{1/8}$.
Neither the \emph{taut-string approximation} nor the
\emph{quasi-static approximation} can be applied, and in contrast to
the other cases the scaling function has to be evaluated numerically.

Eventually, at a time $t_L^\|$, the tension will have propagated
through the whole polymer, i.e.\ $\ell_\|(t_L^\|)\approx L$.
Subsequently, the characteristic longitudinal scale is the contour
length $L$. One would expect that regular perturbation theory would
then become valid.  Surprisingly, for $t_L^\|\gg t_\fex$ the
\emph{Release} scenario provides an exception. The contraction
dynamics exhibits an intermediate regime of \emph{homogeneous tension
  relaxation} (white in Fig.~\ref{fig:tf}).  Its asymptotic power-law
form is revealed by the same \emph{quasi-static approximation}
applicable during the preceding tension propagation, but with the
separation ansatz $f(\bar s,t)\sim h(\bar s)(\zeta L^2\!/\ell_p t
)^{2/3}$ instead of the single-variable scaling ansatz. It solves
Eq.~(\ref{eq:master}) analytically with a roughly parabolic stationary
tension profile $h(\bar s)$. The homogeneous tension relaxation
dominates the short-time relaxation for $t\ll t_\star=(\zeta
L^2/\ell_p)^4$ if $\fex\to\infty$ (i.e.~$t^\|_L \to0$).

To make contact with experiments, it is desirable to derive the
consequences for pertinent observables such as the end-to-end
distance. Integrating the longitudinal Eq.~(\ref{eq:eom2}) over $\bar
s$ and $t$ shows that the growth laws $\Delta \bar R(t)=
2\zeta^{-1}\left|\bar F'(\bar s=0,t)\right|$ are directly related to
the tension profiles discussed above. Tab.~\ref{tab:tf} summarizes our
results for the intermediate asymptotic regimes. Note that $\Delta
\bar R$ is a coarse-grained quantity that does not resolve the
``microscopic'' details below the coarse-graining scale $l$. Near the
polymer ends these give relevant contributions obliterating the
predicted $t^{7/8}$ in experiments
\cite{everaers-etal:99,legoff-etal:2002}.  During homogeneous tension
relaxation $\Delta \bar R\propto t^{1/3}$, which we expect to hold for
strongly stretched polymers even if $L\gg\ell_p$ (e.g.\ DNA), at
variance with earlier predictions \cite{brochard-buguin-de_gennes:99}.
The exponent $1/3$ coincides with that obtained by adiabatic
application of the stationary force-extension relation
\cite{bustamante-marko-siggia-smith:94} to a ``frictionless''
\cite{bohbot-raviv-etal:2004} polymer with attached beads at its ends.
Finally, $\Delta \bar R\approx (\zeta \ell_p)^{-1/2}t^{3/8}$ in
$\ell_p-$\emph{Quenches} for $t\ll t_L^\|$. Interestingly, the tension
propagation/relaxation itself can in some situations be directly
monitored experimentally. In $\ell_p$-\emph{Quenches}, the
(longitudinal) radius of gyration mirrors the characteristic bulk
relaxation $f\propto t^{-1/2}$ of the tension
\cite{hallatschek-frey-kroy:04}. In \emph{Electrophoretic-Pulling},
where $\Delta \bar R\propto t$ for $t\ll t_L^\|$, the force on the
grafted end obeys $\fex\propto\ell_\|$.

In conclusion, we have developed a unified theory of non-equilibrium
tension dynamics in stiff polymers based on the scale separation
between the two dynamic correlation lengths $\ell_\perp$ and
$\ell_\|$.  The recovered known results and our new predictions are
summarized in Figs.~\ref{fig:lf},~\ref{fig:tf} and Tab.~\ref{tab:tf}.
Various dynamic regimes should be well realizable for certain
biopolymers and it is an intriguing question, whether the tension
propagation laws $\ell_\|(t)$ govern mechanical signal transduction
through the cytoskeleton
\cite{shankar-pasquali-morse:2002,gardel-etal:2003}.  Inclusion of
hydrodynamic interactions merely produce logarithmic corrections but
would give rise to more interesting effects for membranes. Other
natural generalizations including the transverse nonlinear response
and more complex force protocols (e.g.~\footnotemark[\value{fnt}]) are
currently also under investigation.

\begin{table}[t]
  \caption{Growth laws for the end-to-end distance $\Delta \bar
    R(t)$ in the intermediate asymptotic regimes marked in
    Fig.~\ref{fig:tf}. OPT and MSPT refer to ``ordinary perturbation
    theory'' and ``multiple-scale perturbation theory'',
    respectively.}\label{tab:tf}
\begin{ruledtabular}
    \begin{tabular}{l|c|c} & {\emph{Release}} & {\emph{Pulling}} \\
    \hline linear MSPT &
    \multicolumn{2}{c}{ $2^{{\frac{5}8}^{\phantom 2}}\!\!\!\Gamma(\frac{15}8)^{-1}
    (\zeta \ell_p)^{-\frac{1}2}\,\fex\, t^{\frac78}$} \quad\mbox{} \\ linear OPT
    \cite{granek:97} & \multicolumn{2}{c}{$2^{-\frac34}
    \Gamma(\frac74)^{-1}(L/\ell_p)\,\fex\, t^{\frac34} $} \\ nonlin.~MSPT
    & $ 3.503\, (\zeta
    \ell_p)^{-\frac12}\,\fex^{\frac14}t^{\frac12}$ & $
    \left(\frac{512}{9\pi}\right)^{\!\frac14} (\zeta
    \ell_p)^{-\frac12} (\fex \, t)^{\frac34}$\\ hom.~relaxation & ${\displaystyle
    2\!\cdot\!54^{\frac23} (L/\zeta \ell_p^2)^{\frac13} t^{\frac13}}$ & --- \\
    nonlin.~OPT & ${ 2^{\frac34}\Gamma(\frac14)^{-1} (L/\ell_p)\,
    t^{\frac14}} $ & ${(2/\pi)^{\frac12}(L/\ell_p)
    (\fex\, t)^{\frac12}}$
\end{tabular}
\end{ruledtabular}
\end{table}



\end{document}